\documentclass{llncs}
\usepackage[utf8]{inputenc}

\usepackage{stmaryrd}

\begin{document}

\title{Note on Optimal Trees for Parallel Hash Functions}

\author{Kevin Atighehchi}
\institute{Aix-Marseille Université, Laboratoire d'Informatique 
Fondamentale de Marseille, case~901, 13288 Marseille cedex 9, France\\
\email{kevin.atighehchi@univ-amu.fr}\\
}

\maketitle

\begin{abstract}
A recent work \cite{AR15} shows how  we can optimize a tree based mode of operation for a rate 1 hash function. 
In particular, an algorithm and a theorem are presented for selecting a good tree topology in order to optimize both 
the running time and the number of processors at each step of the computation. Because this paper deals 
only with trees having their leaves at the same depth, the number of saved computing resources is perfectly optimal only for
this category of trees. 
In this note, we address the more general case and describe a simple algorithm which, starting from such a tree topology, 
reworks it to further reduce the number of processors and the total amount of \emph{work} 
done to hash a message.
\end{abstract}

\section{Introduction}

The present work is in line with what has been done for parallel exponentiation \cite{Sti90,Gat91,AMV88a,LKPC05,WLLC06}. 
We consider hash tree modes using a rate-1 hash (or compression) function, \textit{i.e.} a hash (or compression) 
function which needs $l$ invocations of the
underlying primitive to process a $l$-block message.
Assuming a hash tree of height $h$ and the arity $a_i$ of level $i$ (for $i=1 \ldots h$), we define the parallel 
running time to obtain the root node value as being $\sum_{i=1}^h a_i$. A recent work~\cite{AR15} shows that we can select the good parameters
to construct trees having their leaves at the same depth which minimize both the running time and the number of processors.
The aim of this note is to show that, when considering trees without any structural constraint, we can decrease a little more the amount of used processors to 
obtain an optimal running time. The algorithm we propose decreases as much as possible the amount of processors at each level of the tree, and, as a result,
minimizes the amount of \emph{work}.

The paper is organized in the following way. In Section \ref{terminology}, we give some definitions about trees.
In Section \ref{prev_work}, we recall some of the elements of a previous work~\cite{AR15}. Finally, in Section \ref{algo}, we give an algorithm 
which further decreases both the amount of processors and the amount of \emph{work} done to process a message.

\section{Tree structures}\label{terminology}

Throughout this paper we use the 
convention\footnote{This corresponds to the convention used to describe Merkle trees. The other (less frequent)
convention is to define a node as being 
a $f$-input.
} that a node is the result of
a function called on a data composed of the node's children.
A node value then corresponds to an image by such a function and a children of this node can be either
an other image or a message block.
We call a base level node a node at level $1$ pointing to the leaves representing message data blocks. The leaves (or leaf nodes) 
are then at level~$0$. 
We define the arity of a level in the tree as being the greatest node arity in this level.

A $k$-ary tree is a tree where the nodes are of arity at most $k$. For instance a tree with only one node of arity $k$ is said to be a $k$-ary tree. 
A full $k$-ary tree is a tree where all nodes have exactly $k$ children.
A perfect $k$-ary tree is a full $k$-ary tree where all leaves have the same depth. 

\begin{sloppypar}
We also define other ``refined'' types of tree. 
We say that a tree is of arities $\{a_1,a_2, \ldots, a_n\}$ (we can call it a $\{a_1,a_2, \ldots, a_n\}$-aries tree) if 
it has $n$ levels (not counting level $0$) whose nodes at the first level are
of arity at most $k_1$, nodes at level $2$ are of arity at most $k_2$, and so on. We say that a level $i$ is full if all its nodes (others than leaf nodes) 
are of arity (exactly) $a_i$. 
We say that such a tree is full if all its levels are full. 
As before, we say that such a tree is perfect if it is full and if all the leaves are at the same depth.
\end{sloppypar}

\section{Related work}\label{prev_work}

Let us denote by $l$ the block-length of a message. 
We remark that 
$$\lceil \lceil \cdots \lceil \lceil l/a_1 \rceil / a_2 \rceil \cdots \rceil/ a_i \rceil = \lceil l/(a_1a_2 \cdots a_i) \rceil$$ 
for (strictly) positive integers $(a_j)_{j=1 \ldots i}$.
The problem is to find a tree height $h$ and integer arities $a_1$, $a_2$, ..., $a_h$ 
such that $\sum_{i=1}^h a_i$ is minimized. 
Any solution to the problem must necessarily satisfy the following constraints:
\begin{equation}\label{contraintes}
\prod_{i=1}^h a_i \ge l\ \textrm{and}\ 
\left(\prod_{i=1}^{h} a_i\right)/a_j < l\quad \forall\ j \in \llbracket 1,h \rrbracket.
\end{equation}

A solution to this problem is a multiset of arities. Note that with such a solution, we can construct a tree having exactly $l$ leaves, \emph{i.e.} a tree where the
number of nodes of the first level is exactly $\lceil l/a_1 \rceil$, the number of nodes of the second level is  $\lceil l/(a_1a_2) \rceil$, and so on. 
Among all possible solutions, we would like the one which minimizes both the number of processors and 
the amount of \emph{work}. We recall that the amount of \emph{work}, denoted $W_l$, corresponds to the total amount of computation time to process a message of length $l$. 
For a tree having its leaves at the same depth, it can be evaluated as:
$$W_l=l + \lceil l/a_1 \rceil + \lceil l/(a_1a_2) \rceil + \cdots + \lceil l/(a_1a_2\ldots a_{h-1}) \rceil.$$
We recall the third theorem of \cite{AR15} which selects the good parameters 
for a tree having its leaves at the same depth.

\begin{theorem}\label{min_running_time_max_arities}
For any integer $l \geq 2$ there is an unique ordered multiset $A$
of $h_5$ arities $5$, $h_4$ arities $4$, $h_3$ arities $3$ and $h_2$ arities $2$
such that the corresponding tree covers a message size $l$, has a minimal running time
and has first $h_5$ as large as possible, then $h_4$ as large as possible, and then 
$h_3$ as large as possible.
More precisely, if $i$ is the lowest integer such that $l \leq 3^i < 3l$, 
this ordered multiset is defined according to 11 cases:
\begin{itemize}
 \item Case 1: $|A|=i$, $h_5=0$, $h_4=0$, $h_3=i$, $h_2=0$ if 
 $$\left(l \leq 3^i < \frac{9l}{8}\right) \hbox{ or } \left(i < 2 \hbox{ and } 3^i < \frac{3l}{2}\right);$$
 \item Case 2: $|A|=i$, $h_5=0$, $h_4=1$, $h_3=i-2$, $h_2=1$ if 
 $$\left(\frac{9l}{8} \leq 3^i < \frac{81l}{64} \hbox{ and } i \geq 2\right) \hbox{ or } 
 \left(2 \leq i < 4 \hbox{ and } \frac{81l}{64} < 3^i < \frac{27l}{20}\right);$$
 \item Case 3: $|A|=i-1$, $h_5=0$, $h_4=3$, $h_3=i-4$, $h_2=0$ if 
 $$\left(\frac{81l}{64} \leq 3^i < \frac{27l}{20} \hbox{ and } i \geq 4\right);$$
 \item Case 4: $|A|=i-1$, $h_5=1$, $h_4=1$, $h_3=i-3$, $h_2=0$ if 
 $$\left(\frac{27l}{20} \leq 3^i < \frac{3l}{2} \hbox{ and } i \geq 3\right);$$
 \item Case 5: $|A|=i$, $h_5=0$, $h_4=0$, $h_3=i-1$, $h_2=1$ if 
 $$\left(\frac{3l}{2} \leq 3^i < \frac{27l}{16} \hbox{ and } i \geq 1\right) \hbox{ or } 
 \left(\frac{3l}{2} \leq 3^i < \frac{9l}{5} \hbox{ and } i < 3\right)$$ 
 $$\hbox{ or } \left(\frac{9l}{5} \leq 3^i < \frac{9l}{4} \hbox{ and } i < 2\right);$$
 \item Case 6: $|A|=i-1$, $h_5=0$, $h_4=2$, $h_3=i-3$, $h_2=0$ if 
 $$\left(\frac{27l}{16} \leq 3^i < \frac{9l}{5} \hbox{ and } i \geq 3\right);$$
 \item Case 7: $|A|=i-1$, $h_5=1$, $h_4=0$, $h_3=i-2$, $h_2=0$ if 
 $$\left(\frac{9l}{5} \leq 3^i < \frac{81l}{40} \hbox{ and } i \geq 2\right) \hbox{ or } 
 \left(\frac{81l}{40} \leq 3^i < \frac{9l}{4} \hbox{ and } 2 \leq i \leq 3\right);$$
 \item Case 8: $|A|=i-1$, $h_5=1$, $h_4=1$, $h_3=i-4$, $h_2=1$ if 
 $$\left(\frac{81l}{40} \leq 3^i < \frac{9l}{4} \hbox{ and } i \geq 4\right);$$
 \item Case 9: $|A|=i-1$, $h_5=0$, $h_4=1$, $h_3=i-2$, $h_2=0$ if 
 $$\left(\frac{9l}{4} \leq 3^i < \frac{81l}{32} \hbox{ and } i \geq 2\right) \hbox{ or } 
 \left(\frac{81l}{32} \leq 3^i < 3l \hbox{ and } i = 2\right)$$ 
 $$\hbox{ or } 
 \left(\frac{81l}{32} \leq 3^i < \frac{27l}{10} \hbox{ and } i < 4\right);$$
 \item Case 10: $|A|=i-1$, $h_5=0$, $h_4=2$, $h_3=i-4$, $h_2=1$ if 
 $$\left(\frac{81l}{32} \leq 3^i < \frac{27l}{10} \hbox{ and } i \geq 4\right);$$
 \item Case 11: $|A|=i-1$, $h_5=1$, $h_4=0$, $h_3=i-3$, $h_2=1$ if 
 $$\left(\frac{27l}{10} \leq 3^i < 3l \hbox{ and } i \geq 3\right);$$
\end{itemize}
%
where the number $h_3$ is at least $1$ in the first case and can be $0$ in the other cases. 
\end{theorem}

\section{Optimizing the amount of computing resources}\label{algo}

\subsection{Algorithm}

Given the message size $l$, we first apply Theorem \ref{min_running_time_max_arities} to deduce the list of parameters of the tree, denoted $A$, which is a list of arities. 
The tree topology constructed with these parameters has its leaves at the same depth. Among all the possible trees, this topology is optimal for 
the running time and near-optimal for the number of processors and the amount of \emph{work}. Let suppose that $A$ consists of the arities $\{a_1, a_2, \ldots, a_h\}$, where
$a_1$, $a_2$, ..., $a_h$ are written in decreasing order. Thus, $a_1$ is the arity of the base level (level 1), 
$a_2$ the arity of the second level, and so on, until we reach the root of arity $a_1$. 
Then, we denote by $N_1$, $N_2$, ..., $N_h$ the rightmost nodes of the tree, 
starting from the parent of the rightmost leaf up to the root node, 
and by $r_1$, $r_2$, ..., $r_h$ their arities. We recall that $r_i \leq a_i$ for $i=1 \ldots h-1$, meaning that there is room for optimization on 
the right side of the tree.
Let us now describe an algorithm which will always work on these nodes, placed along the same path.
We repeat the following operations, starting from the root node (with index $i=h$), until we reach a rightmost leaf:
\begin{enumerate}
 \item  We set $i=i-1$. Let $l_i$ be the number of leaves of the subtree rooted at $N_i$. If $l_i=1$, we replace $N_i$ by the single leaf of this subtree (the subtree rooted 
 at $N_i$ is then reduced to a single leaf node) and the algorithm terminates. Otherwise, we go the the following step.
 \item If $N_i$ is of arity exactly $a_i$, then we return to step 1. Otherwise, we seek the largest integer $j$ such that
 \begin{equation}
  \prod_{k=j}^i a_k \geq l_i.
 \end{equation}\label{eq_checking}
 We transform the subtree rooted at $N_i$ accordingly. This subtree is of height $i-j+1$ and has $\lceil l_i/a_j \rceil$ nodes at the base level,
 $\lceil l_i/(a_ja_{j+1}) \rceil$ at the second level, ..., $\lceil l_i/(\prod_{k=j}^{i-1} a_k) \rceil$ at level $i-j$. Then, we go to step~1.
\end{enumerate}

The resulting tree has not necessarily all its leaves at the same depth. We denote by $h'$ the length of the path from the root node to its righmost leaf. If the tree
topology has successfully been reworked, then $h'$ is lower than $h$.
By construction, this path corresponds to the shortest root-to-leaf path. The nodes $N_1$, $N_2$, ..., $N_{h'}$ are placed along this path, while the remaining nodes
$N_{h'+1}$, $N_{h'+2}$, ..., $N_{h}$ are placed along an other (longer) path. This transformed tree 
can be described with two lists $A$ and $B$ where $B$ consists of the arities of $N_{h'}$, $N_{h'-1}$, ..., $N_2$, written in this order.

\paragraph{Example.} Let us take a message of 17 983 blocks. This message has to 
be processed using the first case of Theorem \ref{min_running_time_max_arities}, \emph{i.e.} with a tree of nine levels of arity $3$. 
The arities of the rightmost nodes placed along the same path, 
from the root to the rightmost base level node, are 3, 3, 1, 3, 1, 1, 1, 1, 1. Applying our algorithm above,
we obtain at the third iteration the rightmost arities 3, 3, 3, 1, 1, 1, 1, 1. At the end, the rightmost arities are simply 3, 3, 3.

\subsection{Hints}

The proposed algorithm above is ``brute force'', and any method which could avoid to find the largest $j$ verifying the condition (\ref{eq_checking}) would be welcome.
We would prefer an algorithm in which we look at the arities of the rightmost nodes of a rightmost subtree to decide whether this subtree is updatable or not.

Suppose that, at a given iteration of the algorithm above, we have to decide whether the subtree rooted at $N_i$ can be updated. We propose to analyse the arites of $N_i$
and its rightmost descendants. To do so, we have to define an order to compare two sequences of arities. Let us consider the sequence $(r_i, r_{i-1}, \ldots, r_1)$ 
of arities of $N_i$ and its rightmost (ordered) descendants, and an other sequence of arities $(q_{i'}, q_{i'-1}, \ldots, q_1)$ with $i'\leq i$. We say that
$$(r_i, r_{i-1}, \ldots, r_1) \preceq (q_{i'}, q_{i'-1}, \ldots, q_1)$$
if one of the following conditions is satisfied:
\begin{itemize}
 \item $r_i < q_{i'}$,
 \item or there exists an integer $k < i'$ such that $r_{i-j} = q_{i'-j}$ for $j=0 \ldots k-1$ and  $r_{i-j} < q_{i'-j}$ for $j=k \ldots i'-1$,
 \item or $r_{i-j} = q_{i'-j}$ for $j=0 \ldots i'-1$.
\end{itemize}
If none of these conditions hold, we say that $(r_i, r_{i-1}, \ldots, r_1) \succ (q_{i'}, q_{i'-1}, \ldots, q_1)$.
Some examples are: $(1,2,3,\ldots) \succ (1,1,4)$; $(2,1,4,\ldots) \succ (1,1,4)$; $(3,1,1,9,\ldots) \preceq (3,1,2,4)$; $(5,2,1,2,\ldots) \preceq (5,2,1,2)$.~\\

We analyze the first three cases of Theorem \ref{min_running_time_max_arities} and let the reader deduce the others. 
We suppose that the original tree is of height $h$ and that $i < h$.

\paragraph{Case 1.} The subtree rooted at $N_i$
can be updated if 
$$(r_i, r_{i-1}, \ldots, r_1) \preceq (1)$$
that is, if $N_i$ is of arity $1$. Indeed, suppose $N_i$ is of arity 2. With such choice, the most favourable situation is when the number of leaves of this subtree 
is $3^{i-1}+1$, but $3^{i-1}+1 > 3^{i-1}$.

\paragraph{Case 2.} If $i \geq 5$, the subtree rooted at $N_i$, which can accept until $4 \times 3^{i-1}$ leaves,
can be updated if 
$$(r_i, r_{i-1}, \ldots, r_1) \preceq (1,3,1).$$
Let suppose that $N_{i-2}$ is of arity 2. The most farourable situation is when the number of leaves of the subtree rooted at $N_{i-2}$ is 
$4 \times 3^{i-4}+1$. With this assumption, the number of leaves of the subtree rooted at $N_i$ is $4 \times 2 \times 3^{i-3}+4 \times 3^{i-4} + 1$.
This quantity is greater than $3^{i-1}$, meaning that the arity of $N_{i-2}$ should be lower than $2$. Now, let suppose that $N_{i-2}$ is of arity $1$.
The worst situation is when the number of leaves of the subtree rooted at $N_{i-3}$ is $4 \times 3^{i-5}$. With this assumption, the number of leaves
of $N_i$ is $4 \times 2 \times 3^{i-3} + 4 \times 3^{i-5} \leq 3^{i-1}$. Consequently, this subtree can be transformed by using $i-1$ levels of arity $3$.
If~$i = 4$, the subtree rooted at $N_4$ can be updated if $(r_4, r_3, r_2, r_1) \preceq (1,3,1)$, since $2 \times 3 \times 4 \leq 3^3$. If~$i = 3$,
the subtree rooted at $N_3$ can be updated if $(r_3, r_2, r_1) \preceq (1,3,1)$, since $2 \times 4 + 1 \leq 3^2$. Finally, if~$i = 2$, the subtree rooted 
at $N_2$ can be updated if $(r_2, r_1) \preceq (1,3)$.

\paragraph{Case 3.}If $i \geq 12$, the subtree rooted at $N_i$, which can accept until $4^3 \times 3^{i-3}$ leaves,
can be updated if 
$$(r_i, r_{i-1}, \ldots, r_1) \preceq (1,3,1,3,1,3,1,3,1,3,1,2).$$
We can consider the subtree rooted at $N_{i-12}$. Choosing $N_{i-12}$ as being of arity $3$ lead to a subtree wich is not updatable. However, if this one is
of arity 2, the subtree is updatable whatever is its number of leaves.
We let the reader deduce the cases $i < 12$.

\bibliographystyle{plain}
\bibliography{trees}

\begin{thebibliography}{1}

\bibitem{AMV88a}
Gordon~B. Agnew, Ronald~C. Mullin, and Scott~A. Vanstone.
\newblock Fast exponentiation in \emph{GF(2\({}^{\mbox{n}}\))}.
\newblock In {\em Advances in Cryptology - {EUROCRYPT} '88, Workshop on the
  Theory and Application of of Cryptographic Techniques, Davos, Switzerland,
  May 25-27, 1988, Proceedings}, pages 251--255, 1988.

\bibitem{AR15}
Kevin Atighehchi and Robert Rolland.
\newblock Optimization of tree modes for parallel hash functions.
\newblock {\em CoRR}, abs/1512.05864, 2015.

\bibitem{LKPC05}
Mun{-}Kyu Lee, Yoonjeong Kim, Kunsoo Park, and Yookun Cho.
\newblock Efficient parallel exponentiation in gf(qn) using normal basis
  representations.
\newblock {\em J. Algorithms}, 54(2):205--221, 2005.

\bibitem{Sti90}
Douglas~R. Stinson.
\newblock Some observations on parallel algorithms for fast exponentiation in
  gf(2{\^{}}n).
\newblock {\em {SIAM} J. Comput.}, 19(4):711--717, 1990.

\bibitem{Gat91}
Joachim von~zur Gathen.
\newblock Efficient exponentiation in finite fields (extended abstract).
\newblock In {\em 32nd Annual Symposium on Foundations of Computer Science, San
  Juan, Puerto Rico, 1-4 October 1991}, pages 384--391, 1991.

\bibitem{WLLC06}
Chia{-}Long Wu, Der{-}Chyuan Lou, Jui{-}Chang Lai, and Te{-}Jen Chang.
\newblock Fast parallel exponentiation algorithm for {RSA} public-key
  cryptosystem.
\newblock {\em Informatica, Lith. Acad. Sci.}, 17(3):445--462, 2006.

\end{thebibliography}

\end{document}